\documentclass{aa}  

\usepackage{graphicx}
\usepackage{url}                     % For breaking URLs easily trough lines
\usepackage{rotating}
\usepackage{natbib}
\RequirePackage{fix-cm}
\usepackage{txfonts}
\begin{document}

   \title{Solar cycle 24: an unusual polar field reversal}

%   \subtitle{An unusual polar field reversal}

   \author{P. Janardhan\inst{1}, K. Fujiki\inst{2}, M. Ingale\inst{1}, Susanta Kumar Bisoi\inst{3},
          \and Diptiranjan Rout\inst{4}
          %\fnmsep\thanks{Just to show the usage
          %of the elements in the author field}
          }

   \institute{Physical Research Laboratory, Astronomy \& Astrophysics Division, 
                 Navrangpura, Ahmedabad - 380 009, India.\\
                \email{jerry,mingale@prl.res.in}
                \and
                Institute for Space-Earth Environmental Research, Nagoya, Japan\\
                \email{fujiki@isee.nagoya-u.ac.jp}
                \and
                Key Laboratory of Solar Activity, National Astronomical Observatories,
                Chinese Academy of Sciences, Beijing 100012, China.\\
                \email{susanta@nao.cas.cn}
                \and
                Physical Research Laboratory, Space \& Atmospheric Sciences Division, 
                Navrangpura, Ahmedabad - 380 009, India.\\
                \email{diptir@prl.res.in}
            % \thanks{The university of heaven temporarily does not
              %       accept e-mails}
             }
   
   \date{Received --------; accepted --------------}

% \abstract{}{}{}{}{} 
% 5 {} token are mandatory
 
  \abstract
  {It is well known that the polarity of the Sun's magnetic field 
  reverses or flips around the maximum of each 11 year solar cycle.  
  This is commonly known as polar field reversal and plays a key 
  role in deciding the polar field strength at the end of a cycle, 
  which is crucial for the prediction of the upcoming cycle.}
  % context heading (optional)
  % {} leave it empty if necessary  
   {To investigate solar polar fields during cycle 24, using measurements 
   of solar magnetic fields in the latitude range 
   $55^{\circ}$\,--\,$90^{\circ}$ and $78^{\circ}$\,--\,$90^{\circ}$, to report a prolonged 
   and unusual hemispheric asymmetry in the polar field reversal pattern in solar cycle 24.}
  % aims heading (mandatory)
   {This study was carried out using medium resolution line-of-sight synoptic 
	 magnetograms from the magnetic database of the National Solar Observatory 
	 at Kitt Peak (NSO/KP), USA for the period between February 1975 and October 2017, 
	 covering solar cycles 21\,--\,24 and high-resolution line-of-sight synoptic 
	 magnetograms from the Michaelson Doppler Imager instrument onboard the Solar 
	 Heliospheric Observatory. Synoptic magnetograms using radial measurements from the 
	 Heliospheric Magnetic Imager instrument onboard the Solar Dynamics Observatory, 
	 covering solar cycle 23 and 24, were also used.}
  % methods heading (mandatory)
   {We show that the Southern solar hemisphere unambiguously reversed polarity in mid-2013 
	while the reversal in the field in the Northern solar hemisphere started  
	as early as June 2012, was followed by a sustained period of near-zero field strength 
	lasting until the end of 2014, after which the field began to show a clear rise from its 
	near-zero value.  While this study compliments a similar study carried out using 
	microwave brightness measurements \citep{GoY16} which claimed that the field reversal 
	process in cycle 24 was completed by the end of 2015, our results show that the field 
	reversal in cycle 24 was completed earlier ${\it{i.e.}}$ in late 2014.  Signatures of 
	this unusual field reversal pattern were also clearly identifiable in the solar wind, 
	using our observations of interplanetary scintillation at 327 MHz which supported 
	our magnetic field observations and confirmed that the field reversal process was completed 
	at the end of 2014.}
  % results heading (mandatory)
   {}
  % conclusions heading (optional), leave it empty if necessary 
   {}


   \maketitle
%
%-------------------------------------------------------------------

\section{Introduction} \label{S-Intro}
The reversal in the magnetic field polarity of planetary bodies having a 
global magnetic field is a common phenomenon during which the north 
and south pole of the planet's magnetic field reverses and changes 
orientation.   For example, the Earth shows a field reversal which occurs 
on time scales of millennia.  Unlike the Earth however, the Sun's global 
dipole magnetic field flips or reverses polarity every 11 years around 
the maximum phase of each 11 year solar cycle.  During the reversal, 
the polarity of the solar polar fields in both hemispheres reverses or changes 
to the opposite polarity.  This process was first reported in the epoch making 
paper by \citet{BaB61} after the first measurements of the Sun's polar 
magnetic fields in 1959 \citep{Bab59}.  Since then this phenomena has been 
extensively studied \citep{How72,MFS83,WDM84,FMW98,DWi03,Ben04,DiG04,MuS12} 
using synoptic magnetic charts, ground and space based magnetograms, 
observations of polar coronal holes and polar crown filaments.  

The Sun's polar field at both poles attains its maximum at solar minimum, 
while it reduces and runs through zero during solar cycle maximum 
when the reversal of fields occurs at poles. Polar fields have thus been a good 
indicator of the Sun's polar reversal. The time of solar polar field reversals in 
each hemisphere is identifiable as an unambiguous change in the sign of the 
polar fields or in other words, a clearly identifiable zero crossing of the polar 
fields in both hemispheres. In our earlier reported papers 
\citep{JBG10,JaB11,BiJ14,JaB15a, JaB15b}, we studied solar polar magnetic fields 
during solar cycles 21\,--\,23, using magnetic measurements from the 
{\it{National Solar Observatory}}, Kitt Peak ({\it{NSO/KP}}, USA) 
and the {\it{Michelson Doppler Imager}} on board the 
{\it{Solar and Heliospheric Observatory}} ({\it{SoHO/MDI}}) \citep{DFP95}.    
Polar magnetic fields from the solar photosphere extends out into the corona 
and beyond into the interplanetary medium through regions of the corona known 
as polar coronal holes.  Large polar coronal holes begin to form during the 
late declining phase of a solar cycle, become very prominent at solar minimum 
and are the source of high-speed solar wind streams into the heliosphere 
\citep{KTR73,NoK76,Zir77}.  These large polar coronal holes are usually absent 
during the solar maximum phase when the polar field reversal process is ongoing 
and develop only months after the reversal \citep{KiP09}.  They are, thus, another 
commonly used indicator of polar reversal process. 

The reversal in the current solar cycle 24 has also been studied by many researchers 
\citep{SKa13,KHP14,BiJ14,JaB15a,SuH15,GoY16}.  It is to be noted however, that the 
reported time of polar field reversal process for the current cycle 24 by the different 
researchers is different. Using polar coronal hole area measurements, deduced 
from data of the {\it{Helioseismic and Magnetic Imager}} ({\it{HMI}}) on board 
the {\it{Solar Dynamic Observatory}} ({\it{SDO}}) \citep{LeT12}, the 
epoch of polar field reversal in the Northern hemisphere, in cycle 24, was 
estimated to be mid-2012 \citep{KHP14}.  In another study, \citet{SuH15} reported, 
using high cadence HMI 720s magnetograms, that the reversal of polar fields in the 
Northern hemisphere occurred in November 2012, while the reversal in the Southern 
hemisphere occurred in March 2014.  Using line-of-sight HMI magnetograms 
for a 5 year period starting from April 2010, \citet{PMC15} reported the 
reversal times in the north and south to be April 2014 and Feb 2013, respectively.  
More recently, using 17 GHz microwave images and high latitudes prominence 
activity, \citet{GoY16} reported, that the polar reversal in the Southern hemisphere 
occurred around June 2014, while in the Northern hemisphere the reversal was 
completed only by October 2015. The ambiguous nature of polar field reversal 
in cycle 24 is thus very clear as seen from Table 1, which summarises the 
various estimates of polar field reversal times for the 
Northern and Southern solar hemispheres by different workers. It must be noted 
here that taking into account the different methods used, the measurement 
uncertainties and the difficulty in assigning a given date to a process which 
is slowly changing with time, one can consider the reported times as homogenous 
if the difference is $\leq$ six months.  From Table 1 it is clear that there is
a difference of about a year between our estimate for the end of the reversal in 
the North and that of Gopalswamy (2016).  There is also a difference of about 
a year between the estimates of Pastor Yabar et al., (2015) and Sun et al., (2015) 
for the reversal time in the south.

In the present paper, we have extended our earlier studies 
\citep{JBG10,JaB11,BiJ14,JaB15a, JaB15b} of solar polar fields using magnetic field
\begin{table}[h]
	\centering
		\begin{tabular}{lcc}
		\hline \\
			{\bfseries{Authors}}&{\bf{Reversal Time}}&{\bf{Reversal Time}}\\
			       &{\bf{(North)}}&{\bf{(South)}} \\
			\hline \\
			\citet{KHP14}&Jun. 2012&-\\
			&& \\
			\citet{SuH15}&Nov. 2012&Mar. 2014 \\
			&& \\
			\citet{PMC15}&Apr. 2014&Feb. 2013 \\
			&& \\
			&Oct. 2012& \\
			\citet{GoY16}&to& Jun. 2014\\
			&Sept. 2015 \\~\\
			&Jun. 2012&\\
			This study&to& Nov. 2013 \\
			&Nov. 2014 \\
			\hline \\
		\end{tabular}
	\caption{Estimates of the time of polar field reversals in 
	 solar cycle 24 as reported by different groups of researchers.}
	\label{Table 1}
\end{table}
measurements from {\it{NSO/KP}}, covering solar cycles 21\,--\,24, spanning 
the period from February 1975 to October 2017, to investigate the unusual manner 
in which the solar polar field reversal took place in cycle 24. Synoptic maps 
of solar wind velocities obtained using ground-based interplanetary scintillation 
measurements at 327 MHz were also used in the present study to establish the 
development of polar coronal holes in order to investigate the polar reversal 
pattern in cycle 24.

The onset of the current cycle 24 was delayed, due to the 
extended and deep solar minimum experienced in cycle 23, with 
the first sunspots of the cycle 24 appearing only in March 2010 
instead of the expected time of December 2008 \citep{JRL11}.   
Also, solar cycle 24 is the fourth in a series of 
successively weaker cycles since cycle 21, and is actually the 
weakest of the four.  Solar cycle 23, not only experienced one 
of the deepest minima in the past 100 years but the peak 
smoothed sunspot number (V2.0) was $\sim$116 in April 2014, 
making it the weakest sunspot cycle since cycle 14, which 
had a smoothed sunspot number peak (V2.0) of $\sim$107 in 
February 1906.  It is known that the polar field strength at the end 
of a given cycle is determined by the amount of flux cancelled during 
polar field reversal, which is predominantly decided by the systematic 
tilt angle distribution of bipolar sunspot regions, in combination with 
processes of solar differential rotation and advection 
\citep{BaB61,Lei69,WNS89,Pet12}.  It is, therefore, imperative 
that we try and understand how the complex nature 
of solar activity in cycle 24 has contributed to the solar polar field 
reversal process during cycle 24.
%
%----------------------------- Begin Fig 1 ---------------------------
%
\begin{center}
	\protect\begin{figure*}[!ht]
			\vspace{17.6 cm}
		\includegraphics{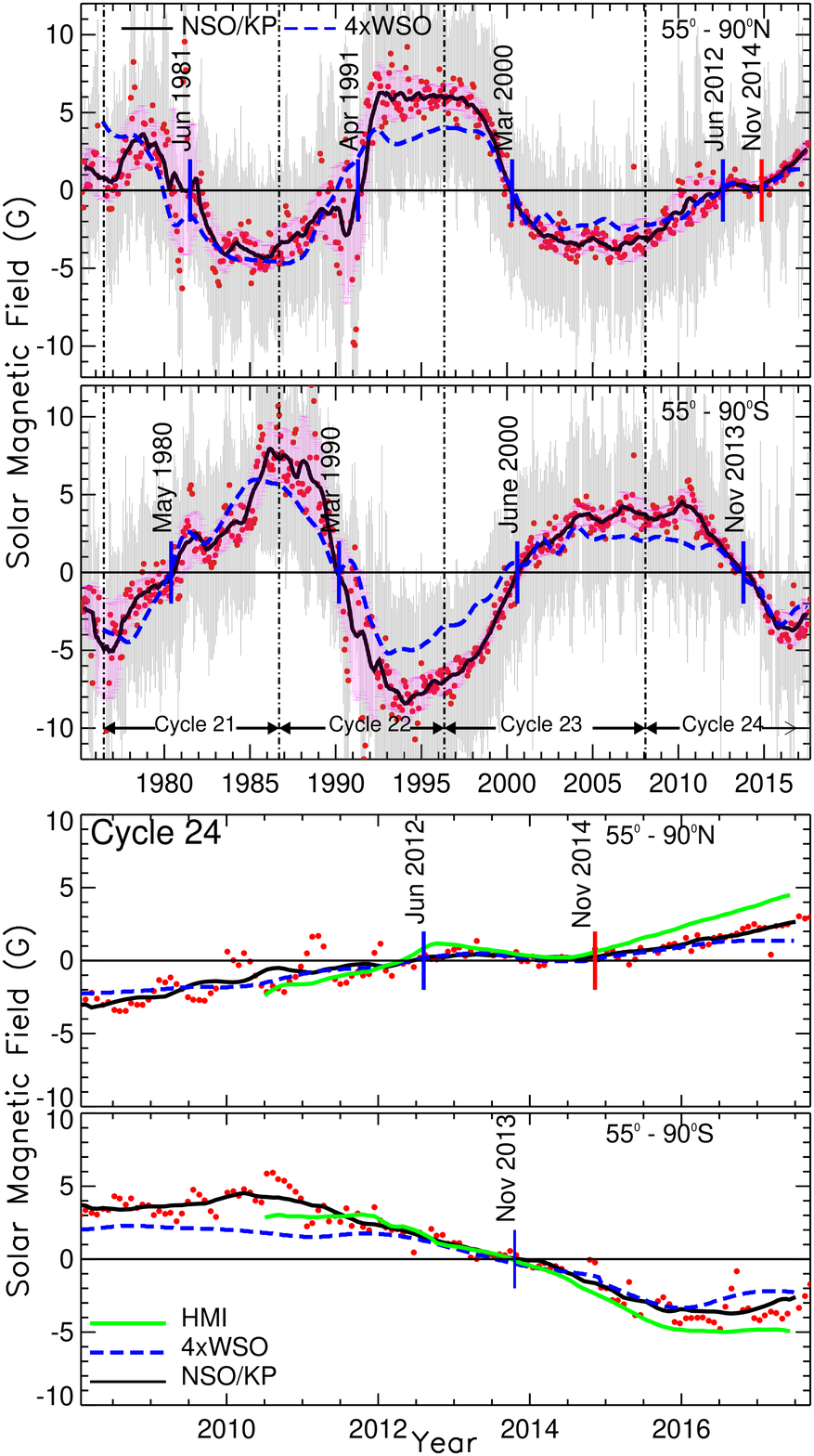}
		\caption{The upper two panels show, by the filled dots (red) with 1 $\sigma$ 
		error bars (grey), the signed values of polar field strengths from NSO/KP, in 
		the latitude range $55^{\circ}$\,--\,$90^{\circ}$, in the solar Northern (top) and 
		Southern (bottom) hemispheres spanning period covering solar cycles 21\,--\,24, 
		each of which is demarcated by vertical dashed lines.   A smoothed solid black 
		curve with 1 $\sigma$ error bars (pink) is drawn through the data points.  
		Also, overplotted by a dashed blue curve is the signed values of polar field 
		strengths from WSO. In order to eliminate yearly geometric projection effects 
		at the poles due to the rotation of the Earth on the measured line-of-sight 
		fields, a 1.58 year (or 20nHz) low pass filter was used to filter WSO 
		polar fields.  The time of reversal of polarity in each hemisphere, determined 
		from NSO/KP smoothed curve, is marked by small blue solid lines in each panel, 
		while the time of completion of polar reversal in the solar north is indicated 
		by a small red solid line in cycle 24. The lower two panels show only cycle 24, 
		and to avoid clutter and enhance clarity, the error bars have been left out.
		Overplotted by a solid green curve, in the lower two panels, is the signed 
		values of polar field strengths from SDO/HMI. In cycle 24, the Southern 
		hemisphere underwent a clean unambiguous reversal in November 2013, while the 
		Northern hemisphere has shown an extended zero-field condition after the 
		first reversal in June 2012.  The reversal in the Northern hemisphere was 
		completed only by November 2014.}
		\label{fig1}
	\end{figure*}
\end{center}
%
%----------------------------------- End Fig 1 ------------------------
%

\section{Data and Methodology} \label{sec:dat}
For computation of photospheric magnetic fields at different latitude ranges, 
we used medium-resolution line-of-sight synoptic magnetograms from both 
the {\it{NSO/KP}} database and the {\it{Synoptic Optical Long-term Investigation 
of the Sun facility}} ({\it{NSO/ SOLIS}}).   The synoptic magnetogram used 
were from February, 1975 to October, 2017, covering Carrington rotation (CR) 
CR1625\,--\,CR2195.  Each synoptic map is generally made from several full-disk 
daily solar magnetograms observed over a Carrington rotation period covering 
27.2753 days. These maps, available online as standard FITS files, contain 
photospheric magnetic fields in units of Gauss, in sine of latitude and longitude 
format of 180 $\times$ 360 pixels.  Also, for a comparison of photospheric 
fields, obtained from medium-resolution {\it{NSO/KP}} synoptic magnetograms, 
we computed photospheric fields in the corresponding latitude ranges for cycle 24, 
using high-resolution radial synoptic magnetograms from the {\it{SDO/HMI}}.  
These maps have 1440 $\times$ 3600 pixels in sine of latitude 
and longitude format, covering CR2096\,--\,CR2194 (from April 2010 to Aug 2017).  
For cycle 23, we used high-resolution line-of-sight synoptic magnetograms 
from the {\it{SoHO/MDI}}, available as maps having 1440 $\times$ 3600 pixels 
in sine of latitude and longitude format, covering CR 1911\,--\,CR2080 
(from April 1996 to February 2009). In order to compare the results from 
{\it{NSO/KP}}, we degraded (by averaging) the resolution of both the 
{\it{SoHO/MDI}} and {\it{SDO/HMI}} to the resolution of {\it{NSO/KP}}.  
In addition, we also used line-of-sight measurements of 10-day averaged 
values of polar fields obtained at the {\it{Wilcox Solar Observatory}} 
({\it{WSO}}) in the latitude range poleward of $55^{\circ}$, covering 
solar cycles 21\,--\,24.  It is to be noted however, that due to the inclination 
of the Earth's orbit to the Sun's equator, one of the poles will not be clearly 
visible.  The data at {\it{NSO/KP}} and {\it{SOHO/MDI}} 
is made available online, to the general user, after having been corrected 
for this effect, while the {\it{SDO/HMI}} data is available 
without this correction being incorporated.  We therefore computed polar fields 
for {\it{SDO/HMI}} in the latitude range $55^{\circ}$\,--\,$78^{\circ}$ leaving out the 
pole-most regions.  Additionally, in order to eliminate yearly geometric projection 
effects at the poles due to the rotation of Earth on the measured line-of-sight fields, 
we used a 1.58 year low pass filter to filter WSO polar fields.  Thus, the obtained 
filtered polar fields are free from the effect of any kind of annual periodicities, 
which if not removed may result in showing multiple zero-crossing of the polar fields 
during solar polar field reversal thereby causing confusion about the occurrence of 
multiple polar field reversals.  
%
%----------------------------- Begin Fig 2 -----------------------------
\begin{center}
	\protect\begin{figure}[!ht]
		\vspace{16.2cm}
		\includegraphics{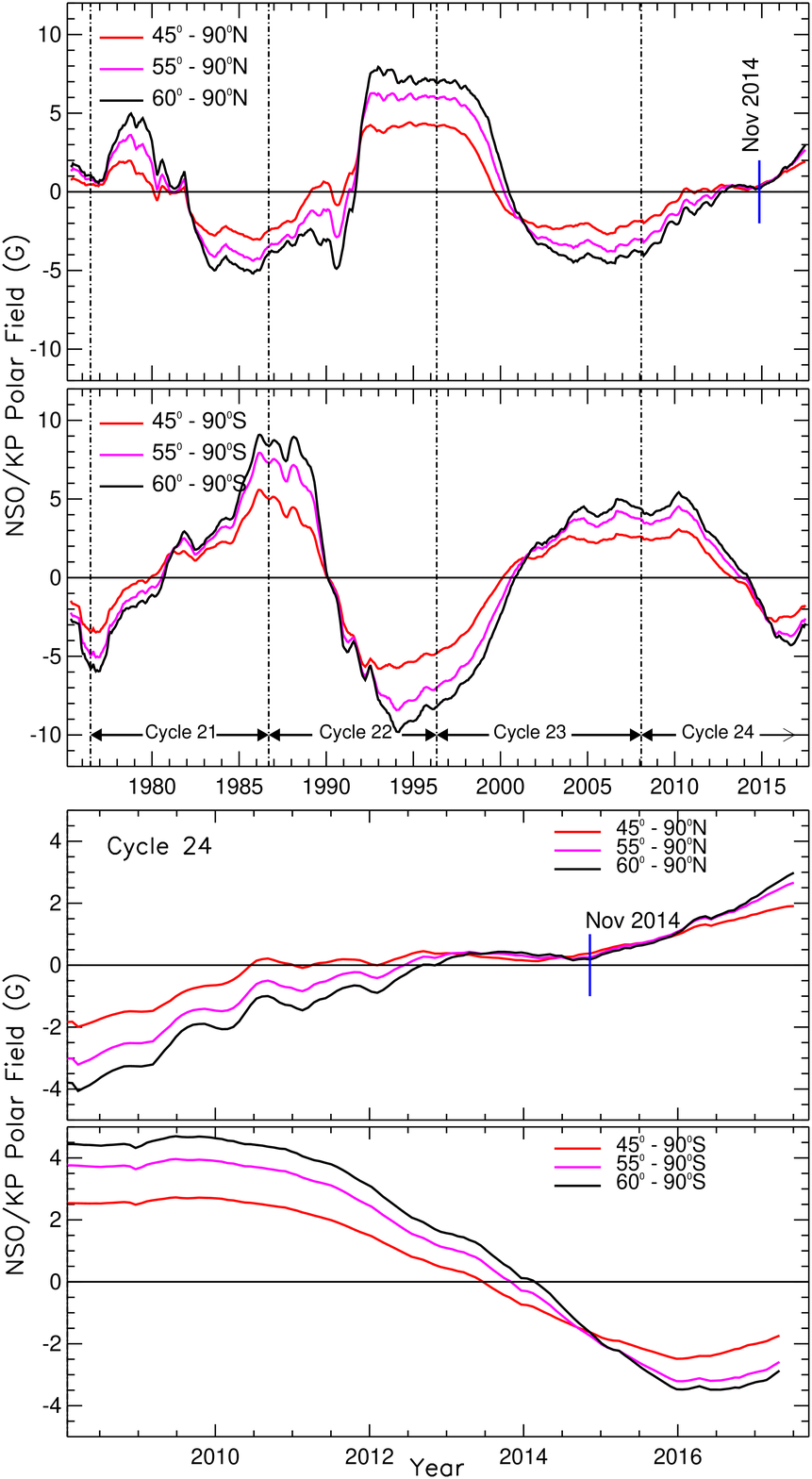}
		\caption{The upper panels show a comparison of the signed values of polar field 
		strengths, obtained from line-of-sight NSO/KP magnetograms, in the latitude 
		ranges, $45^{\circ}$\,--\,$90^{\circ}$, $55^{\circ}$\,--\,$90^{\circ}$, and 
		$60^{\circ}$\,--\,$90^{\circ}$ 
		for the Northern (top) and Southern hemisphere (bottom), for the period 
		February 1975\,--\,November 2016, covering solar cycles 21\,--\,24.  The reversal of 
		polar fields is clearly seen from the zero-crossing of fields around the solar maximum 
		phase of each cycle. The lower panels show a blow up portion for cycle 24. The extended 
		zero-field conditions of polar reversal in the Northern hemisphere is 
		apparent at all latitude ranges, which is completed only around 
		November 2014 as indicated by a small blue solid line in cycle 24.}
		\label{fig2} 
	\end{figure}
\end{center}
%----------------------------------- End Fig 2 -----------------------
% 

For each synoptic map, photospheric magnetic fields were first estimated for each 
of the 180 arrays in sine of latitude by taking a longitudinal average of 
the entire 360 array of Carrington longitude. Thus, each CR map, previously in 
the form of a 180 $\times$ 360 array, was reduced to an array of 180 $\times$ 1. 
Thereafter, for computing magnetic fields at selected latitude ranges, we averaged the 
longitudinal strip of 180 $\times$ 1 over appropriate latitude ranges.  For the present 
study, we selected three latitude zones: toroidal fields, polar fields, and polar cap fields 
in the latitude range $0^{\circ}$\,--\,$45^{\circ}$, $55^{\circ}$\,--\,$90^{\circ}$, and 
$78^{\circ}$\,--\,$90^{\circ}$, respectively.  It is, however, to be noted that the latitudinal 
range defined as polar has been arbitrary with different researchers using different 
latitude ranges for estimation of polar fields such as poleward of $45^{\circ}$ 
\citep{JaB11,BiJ14}, $55^{\circ}$ (Wilcox Solar Observatory polar fields, 
\url{http://wso.stanford.edu/Polar.html}, $60^{\circ}$ \citep{Tom11,GoY12,SuH15,GoY16}, 
and $70^{\circ}$ \citep{MuS12}.  For the present study, in order to compute polar fields, 
we preferred the latitude range between $55^{\circ}$ and $90^{\circ}$, which is same as the 
latitude range used to measure polar fields at WSO.  Similar latitude ranges ($>$ $50^{\circ}$)  
have been used by \citet{KHP14} in order to compute polar coronal hole area to report 
the polar reversal process in cycle 24.  \citet{Alt11} reported the appearance of 
the Fe XIV emission feature in the corona at latitudes $>$ $50^{\circ}$, prior to 
the solar cycle maximum, which started to drift to the poles over time and which he 
called as rush-to-the-poles.  The author suggested that the solar maximum usually 
occurs when a linear fit to the rush-to-the-poles data reaches $76^{\circ}$ $\pm$ $2^{\circ}$.  
We, therefore, also computed polar cap fields in the latitude range $78^{\circ}$-$90^{\circ}$ 
to study the polar field reversal process at the polemost latitudes.  Earlier, \citet{Ben04} 
used {\it{SDO/HMI}} data in the latitude range poleward of $78^{\circ}$ to report the 
behaviour of polar fields during polar field reversal in cycle 23. In our earlier work 
\citep{JBG10}, we used also the same latitude range to report the correlation of meridional 
flow speed with polar fields during cycle 23.

For studying solar wind signatures due to outflows from polar coronal hole 
regions, we constructed synoptic velocity maps using 327 MHz interplanetary 
scintillation (IPS) observations from the three station IPS 
facility of the Institute for Space-Earth Environmental Research (ISEE), 
Nagoya University, Nagoya, Japan.  IPS is the method of studying random 
temporal variations of the signal intensity of distance extra-galactic
radio sources using ground based telescopes, typically observed at 
meter wavelengths \citep{HSW64,JAl93,ABJ95,JaB96,MoA00,BiJ14b}.  A tomographic 
technique \citep{KoT98} was used to produce synoptic solar wind
velocity maps.  Polar coronal holes, as the name suggests, are identifiable 
in the polar regions of the synoptic maps as large, extended regions of high 
velocity solar wind outflows, appearing after polar field reversal.
%
%----------------------------- Begin Fig 3 ---------------------------
\begin{center}
	\protect\begin{figure*}[!ht]
				\vspace{13.0cm}
				\includegraphics{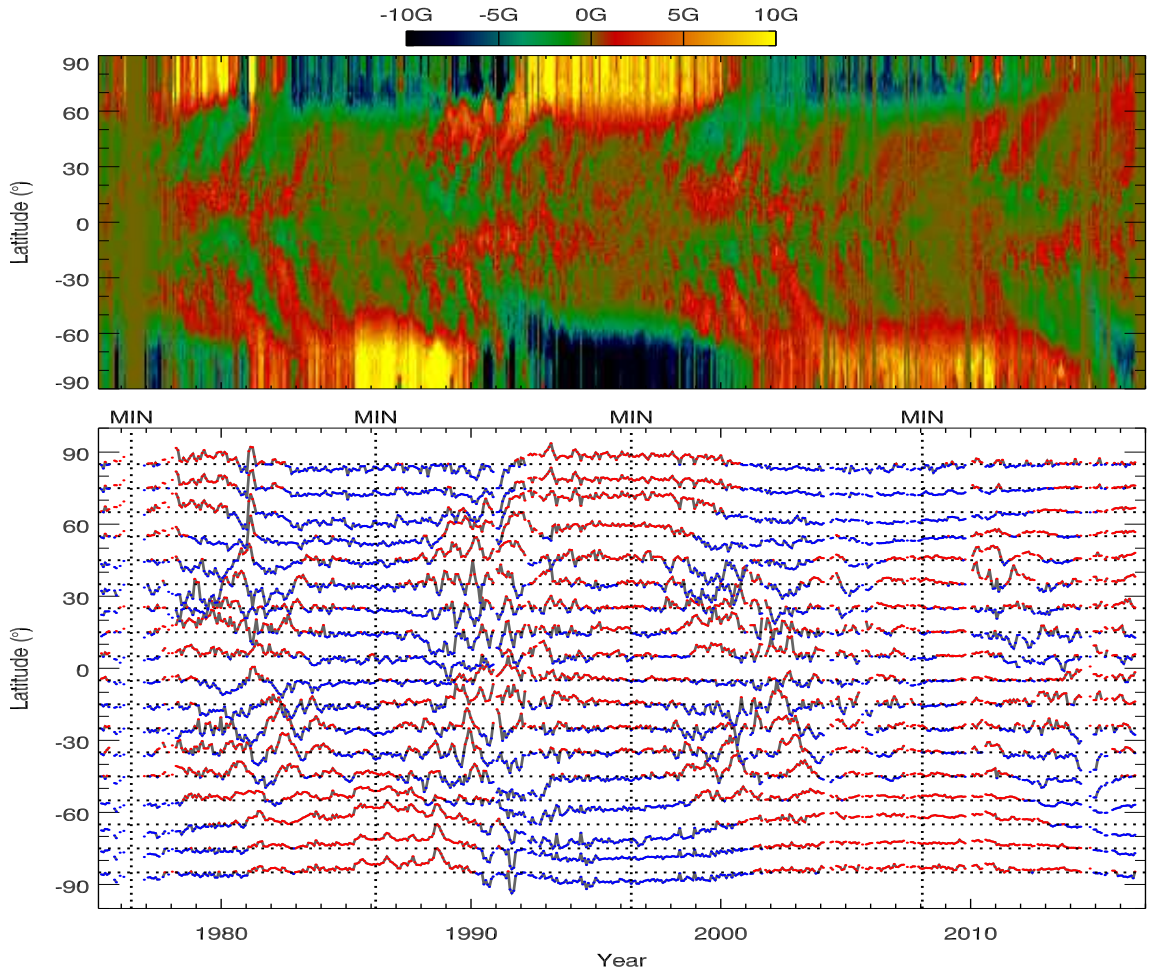}
		\caption{A zonally averaged magnetic diagram (upper) produced 
			after re-gridding individual synoptic magnetograms from 
			sine latitude to latitude with a spatial resolution of 
			one degree and then taking longitudinal average of the 
			magnetic field in each latitude by using only the data 
			points with weak magnetic field strength 
			($\left| B\right| <$ 10 G) to suppress the effects of 
			sunspots in the mid-latitudes and to reproduce the 
			{\bf{magnetic flux transport to the poles}} with higher 
			contrast.  The lower panel displays the temporal profiles 
			of the mean magnetic field in every 10 degree latitude bin 
			obtained from the butterfly diagram in the upper panel, 
			with the red and blue points denoting positive and negative 
			polarity, respectively.}
		\label{fig3}
	\end{figure*}
\end{center}
%----------------------------------- End Fig 3 -----------------------
%   
\section{Results and Discussions}
\subsection{Photospheric signatures}
The upper panels in Figure \ref{fig1} show, by red filled dots with 1 $\sigma$ error bars, 
the signed value of the variation of polar fields computed in the latitude range 
$55^{\circ}$\,--\,$90^{\circ}$, in the Northern (top) and Southern (bottom) hemispheres, 
during solar cycles 21\,--\,24.  Each cycle is demarcated by a vertical dashed line.  
A smoothed curve through the data points in both hemispheres is shown by a solid black 
line with 1 $\sigma$ error bars (pink), after using a 1.58 year filter so as to
remove projection effects caused by the rotation of the Earth.  The 
time of solar polar field reversals in each hemisphere, identifiable as a change in 
the sign of the field {\it{i.e.}} a clearly identifiable zero crossing of the field, 
is shown by small vertically oriented solid blue lines from cycles 21\,--\,24 and 
labelled with the month and year of the reversal.  The lower two panels show a blown 
up portion of Fig.\ref{fig1} only for solar cycle 24.  For comparison, shown in blue are
the filtered polar fields from {\it{WSO}} using the 1.58 year filter and the 
signed values of polar field strengths from SDO/HMI, in green, in the lower two panels of 
Fig.\ref{fig1}.  As mentioned earlier, it is to be noted that polar fields at 
{\it{WSO}} are usually measured in the latitude range $55^{\circ}$\,--\,$90^{\circ}$.  It 
is evident from the variation of fields as shown by the smoothed curves of {\it{NSO/KP}} 
and {\it{WSO}} that their behaviour is similar.  Thus, medium-resolution 
NSO/KP magnetograms are still useful in studying the nature of the large scale polar 
fields.  It can also be seen from Fig.\ref{fig1} that the time of solar polar field 
reversals in each solar hemisphere, during cycle 21\,--\,24, for {\it{NSO/KP}} and 
{\it{WSO}} are both around the same time.  Also, overplotted by a solid curve in 
green in the lower panels of Fig.\ref{fig1} are the polar fields from {\it{SDO/HMI}}, 
in the latitude range $55^{\circ}$\,--\,$78^{\circ}$, for cycle 24. It can be seen that 
the nature of polar fields and the time of reversal in cycle 24, as shown by smoothed 
curves in black, blue, and green in the lower two panels, are by and large similar 
for the {\it{NSO/KP}}, {\it{WSO}} and the {\it{SDO/HMI}} data.  The error 
bars have been omitted in the lower two panels to reduce clutter and enhance clarity.

The hemispheric asymmetry in polar field reversal is clearly evident from 
Fig.\ref{fig1} for Cycles 21\,--\,24.  However, unlike the previous cycle 23, 
the polar field reversal in cycle 24 shows an unusual nature.  It can be seen 
from Fig. \ref{fig1} that the Southern hemisphere underwent a clean, clearly 
identifiable field reversal in Nov 2013.  The reversal in the Northern hemisphere, 
on the other hand, started in June 2012 and continued showing an extended near-zero 
field condition until November 2014, indicating that the completion of reversal 
of polar fields in the north in cycle 24 had not yet taken place.  The 
fields in the Northern hemisphere showed a rise after November 2014, indicating 
the completion of polar field reversal at that time.  The WSO polar fields also 
show two clearly identifiable changes in the sign of the field for the Northern 
hemisphere in cycle 24 with the first or initial reversal starting as early as in 
June 2012 and the second change in the sign of the field occurring almost 2 
years later in November 2014, indicating the completion of the field reversal at 
that time.  Such a field reversal pattern, separated by two years 
between the two hemispheres and the prolonged zero-field conditions delaying 
the completion of reversal in one of the hemispheres by more than two years is 
very unusual, as field reversals usually occur in both the hemispheres within six 
months to a year of each other, and thereby completing the reversal process.  This 
pattern, as observed in cycle 24, is clearly unusual as is apparent 
in Fig.\ref{fig1} from field reversals seen in cycles 21, 22 and 23.  

Other researchers have also reported similar hemispheric differences in polar 
reversal in cycle 24. As mentioned earlier, it is 
%
%----------------------------- Begin Fig 4 ---------------------------
\begin{center}
	\protect\begin{figure}[!ht]
				\vspace{12.3cm}
				\includegraphics{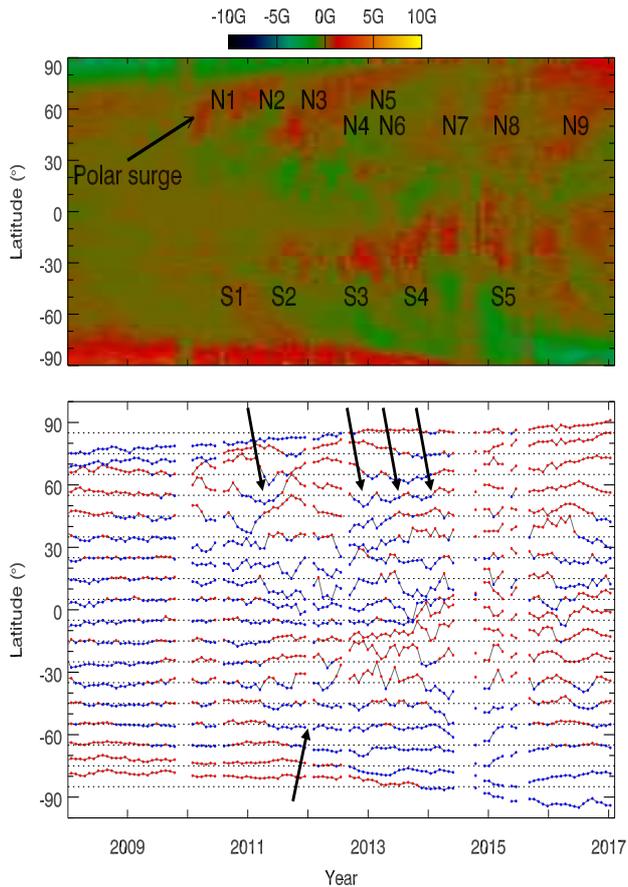}
		\caption{The zonally averaged magnetic diagram (upper panel) and the temporal 
		profiles of mean magnetic fields (lower panel) same as shown in Fig.\ref{fig3}, 
		but only for Cycle 24.  The zonally averaged magnetic diagram is shown with 
		magnetic field values saturated at 10 G. The polar surges in the 
		north are numbered from N1 to N9, and that in the south are numbered from S1 to S5. 
		The polarity changes in mean magnetic field profiles in the north and the south 
		are indicated by arrows. Multiple arrows in the north indicate multiple change 
		in polarity in the north compared to a single clean change in polarity in the south.}
	\label{fig4}
	\end{figure}
\end{center}
%----------------------------------- End Fig 4 -----------------------
% 
to be noted that \citet{KHP14} 
reported the start of the Northern hemisphere reversal in mid-2012 using polar 
coronal hole areas and EUV emission in the corona above latitudes of $50^{\circ}$.
Using polar fields computed from {\it{HMI}} data poleward of 
60${^{\circ}}$, \citet{SuH15} reported a reversal in the north in late 2012 
and a reversal in the south in early 2014.  The use of different latitude ranges can 
result in the difference of the time of reversal of polar fields, which is 
evident from Figure \ref{fig2}. The upper panels in Fig.\ref{fig2} show the 
variation of polar fields computed in the latitude range poleward of 
$45^{\circ}$, $55^{\circ}$ and $60^{\circ}$, for the Northern and Southern 
hemispheres.  It is clear from the upper panels of Fig.\ref{fig2} that the 
overall behaviour of polar fields is the same whether one considers the 
latitude range poleward of  $45^{\circ}$, $55^{\circ}$ or $60^{\circ}$.  
The only difference that one sees is the different times of zero-crossing 
of the field during polar field reversal in each cycle.  It is however to 
be noted that zero-crossing of the field usually occurs earlier for 
the fields at the lower latitude than the higher latitude.  This can be understood 
from the fact that the cancellation of the opposite fluxes usually occurs first at 
lower latitudes.  So the zonally averaged polar fields estimated over different 
latitude  ranges will have different times for polar reversal.   Surprisingly 
though, as evident from the lower panels of Fig.\ref{fig2}, the recovery of polar 
fields for all  the latitude ranges, from the prolonged zero-field condition in 
the Northern hemisphere in cycle 24, started nearly around the same period, in 
November 2014, as indicated by a small blue vertical line in the lower panel 
of Fig. \ref{fig2}. 

In an another study of polar field conditions in cycle 24, using 17GHz microwave 
images obtained by the Nobeyama Radio Heliograph (NoRH) \citet{GoY16} stated that 
the extended zero field condition in the solar Northern hemisphere lasted until 
October 2015.  The authors showed that there was an absence of microwave brightness 
enhancement at the Northern polar region for three years following 2012, 
indicating a zero field condition in the Northern hemisphere until 2015.  It was 
also shown \citep{GoY16} that during this zero field period for three years, i.e. 
until 2015 the microwave brightness enhancement showed undulating signatures which 
were interpreted as polar surges of alternating polarities reaching the north polar 
region.  The polar surges are magnetic flux channels which can be best viewed in 
a magnetic butterfly diagram. We therefore constructed a magnetic butterfly diagram, 
for the period from February 1975 to December 2016, using {\it{NSO/KP}} magnetic 
field observations, in order to investigate the polar surges. 

Figure \ref{fig3} (top) shows such a zonally averaged magnetic field butterfly
diagram constructed after removing the strong magnetic field regions, typically 
associated with sunspot regions in the equatorial belt in both hemispheres, up to
about 35${^{\circ}}$ in latitude.  As stated earlier,  the original synoptic 
magnetogram data is a two dimensional map with 360 pixels in longitude and 180 pixels in 
Sine latitude.  Each map was first re-gridded from sine latitude to latitude with 
a spatial resolution of one degree.  A longitudinal average of the magnetic field 
in each latitude was then taken by considering only the data points with weak 
magnetic field strength ($\left| B\right| \leq$ 10 G).  The strong magnetic fields 
associated with active regions and sunspots were therefore suppressed.  The resulting 
magnetic butterfly diagram reproduces the polar surges with much more clarity and with 
higher contrast.  Fig. \ref{fig3} (bottom) displays temporal profiles of the mean 
magnetic field in every 10${^{\circ}}$ bin in latitude obtained from the butterfly 
diagram in the upper panel.  The red and blue points on the profiles (in grey), denote 
positive and negative polarity, respectively.  A reversal of the polar field will be 
seen as a change in colour from red to blue and vice-versa, which is apparent in solar 
cycles 21--23.   However, the polar reversal in cycle 24 shows an unusual pattern.  The 
polarity reversal in cycle 24 can be noticed from a change of the polarity of from blue 
(red) points to red (blue) points, respectively, in the Northern (Southern) polar 
region above the latitude $>$ 55${^{\circ}}$.  

For an enhanced view, Figure \ref{fig4} shows the zonally averaged magnetic butterfly 
diagram (upper panel) and temporal profiles of mean magnetic fields (lower panel) 
only for cycle 24. As against the normal pattern of multiple surges of one polarity, 
it is evident from the butterfly diagram in Fig.\ref{fig4} that there are  
multiple instances of surge activity of both polarities (positive and negative) 
in the Northern hemisphere in cycle 24 while the polar surges in the Southern 
hemisphere in cycle 24 are of same polarity (negative only). This is also 
clearly established from the profiles in Fig. \ref{fig4} (bottom) which depict 
multiple changes of polarity (red to blue and blue to red) in the Northern 
hemisphere  at latitudes poleward of $>$ 55${^{\circ}}$, while the Southern 
hemisphere showed a clean reversal of polarity from positive (red) to negative 
(blue).  The first poleward surge (of positive polarity) in the north, N1, 
was seen sometime around 2010, when the reversal in the north had begun changing 
polarity to positive.  About a year later, the second poleward surge, N2 
(of negative polarity)
% 
%----------------------------- Begin Fig 5-----------------------------
\begin{center}
	\protect\begin{figure}[!ht]
	       \vspace{11.8cm}
	       \includegraphics{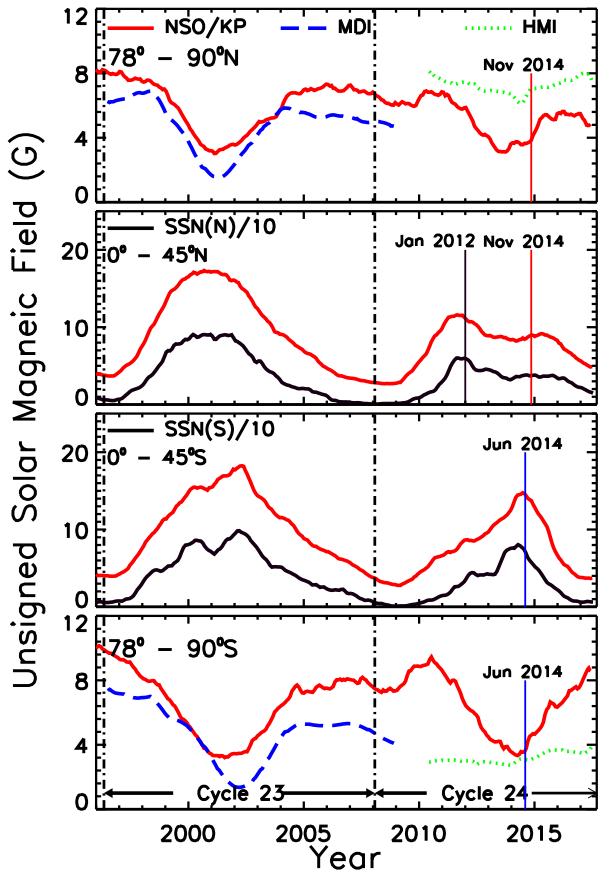}
		\caption{The panels, from top to bottom, show the unsigned values of 
		solar magnetic fields, obtained from NSO/KP (solid curve in red), 
		SoHO/MDI (dashed curve in blue), and HMI (dotted curve in green) 
		magnetograms, in the latitude ranges, poleward of $78^{\circ}$ in 
		the north, $0^{\circ}$\,--\,$45^{\circ}$ in the north, 
		$0^{\circ}$\,--\,$45^{\circ}$ in the south, and poleward of $78^{\circ}$ 
		in the south, respectively, for solar cycles 23\,--\,24. The hemispheric 
		smoothed sunspot number is shown in solid black curve overplotted in 
		the second (north) and third (south) panels. The hemispheric asymmetry 
		in solar activity in cycle 24 is evident with the south showing one 
		activity peak (Jun 2014), while the north showing double activity peaks 
		(Jan 2012 and Nov 2014).  It is also clear from the first and fourth 
		panels that the polar reversal process in the north in Cycle 24 is slow 
		and seems to be completed in Nov 2014, while in the south, it is faster 
		and has been completed in early 2014. The occurrence and completion of 
		polar reversal process happening around the solar cycle maximum is 
		evident from the figure.}
		\label{fig5}
	\end{figure}
\end{center}
%----------------------------------- End Fig 5 -----------------------
%
appeared, reversing the polarity to negative again. This pattern has 
repeated with subsequent surges N3, N4, N5 and N6 of alternate polarity until the 
largest and continuing polar surge N7 of positive polarity appearing in 
mid-2012, when finally the reversal in the north had occurred.  However, 
the re-occurrence of alternate polarity kept the zero-field condition 
in the north until late 2014, when a large surge of magnetic flux, N8 
finally completed the reversal in the north.  The change of polarity in 
the Southern hemisphere is comparatively very clean and was completed in
late 2013.  Multiple large surge activity in the south, indicated 
by S1, S2, S3, S4, and S5 of same polarity except for S1, are clearly 
noticeable in cycle 24 which establishes the clean reversal in the Southern 
hemisphere.  It is, thus, clear from Fig.\ref{fig4} 
that the multiple surges of alternate polarity contributed to the prolonged zero-field 
conditions and the delayed field reversal in the Northern hemisphere.  As stated 
earlier, \citet{GoY16} reported that the extended zero field condition in the Northern 
hemisphere in cycle 24 lasted until October 2015, by which time the field reversal process 
in cycle 24 was complete.  Our magnetic field observations, on the other hand, show 
that the prolonged zero field condition in the Northern hemisphere in cycle 24 existed 
only until late 2014 when the field reversal process was finally completed in the north. 
The delay in completion of polar reversal can be understood from the arrival of polar 
surges of alternate polarity in the polar cap regions until late 2014. 

\subsection{Polar cap fields and meridional flow speeds}

From Fig.\ref{fig4}, it is apparent that, in cycle 24, polar surges 
or tongues of magnetic flux channels were seen to be extended up to 
the polar cap latitudes of  $\sim$75${^{\circ}}$.  It is thus expected, 
that the poleward transport of magnetic flux in cycle 24 can be 
extended all the way to the poles. In fact, it is not uncommon 
as we have seen this during cycle 22 and 23 when such flux transport 
reached up to the latitudes of $\sim$60${^{\circ}}$ \citep{DiG10}. In our 
earlier work \citep{JBG10}, we used polar cap fields in the latitude range 
between $78^{\circ}$-$90^{\circ}$ and showed a very good correlation between 
polar cap fields with the magnetic flux transport for cycle 23.  It is, 
therefore, necessary to verify the polar field variations in the polar 
cap regions beyond $\sim$ 75${^{\circ}}$.  Figure \ref{fig5} shows the 
unsigned values of polar cap field strengths, in the latitude range 
$78^{\circ}$-$90^{\circ}$, obtained from {\it{NSO/KP}}, for the Northern 
(first panel) and Southern (fourth panel) hemispheres, covering solar 
cycles 23 and 24.  For comparison, the polar cap field strengths obtained 
from high-resolution data from {\it{SoHO/MDI}} (dashed blue curve) are 
overplotted at latitudes poleward of $78^{\circ}$ for cycle 23 and from 
{\it{HMI}} (dotted green curve) in the latitude range $60^{\circ}$-$78^{\circ}$ 
for cycle 24, respectively.  It is evident from Fig.\ref{fig5} that the 
polar field strength in the north reached its minimum 
values much earlier than the south showing the hemispheric phase shifts in polar 
field reversal process in cycle 24.  The polar fields after running through the 
minimum value recovers slowly in the Northern hemisphere and the recovery 
of minimum condition has been completed only in late 2014, as indicated by a 
small red vertical line.  Though the polar cap fields in the Southern hemisphere 
showed a faster recovery from minimum condition indicating the completion 
of reversal in the south occurring sometime in mid-2014. The time of completion of 
field reversal in both the hemispheres is around the similar period as 
the change of polarity at the polar cap latitudes from magnetic butterfly diagram 
shown in Fig.\ref{fig4}.

The second panel (Northern hemisphere) and third panel (Southern hemisphere) of 
Fig.\ref{fig5}, shown by a solid red curve, shows the variation of toroidal fields or 
photospheric fields in the latitude range $0^{\circ}$-$45^{\circ}$, derived 
from {\it{NSO/KP}} data, confined to the sunspot activity belt area.  The hemispheric 
smoothed sunspot number (SSN) is overplotted in both the panels shown by a 
solid black curve.  The hemispheric phase shifts in solar activity peak is evident 
from the toroidal fields for both cycle 23 and 24. In cycle 23, a double peak solar 
maximum is apparent for both the hemispheres, which is not the case in cycle 24. The 
double peak maximum is seen only in the Northern hemisphere, while the Southern 
hemisphere peaks only once in June 2014.  It may noted that the reversal in the 
south has been completed during the same period as seen in the fourth panel of 
Fig.\ref{fig5}. The first peak in the north occurred $\sim$2 years earlier than 
in the south. However, the second peak occurred in Nov 2014 when the reversal 
in the north was finally completed as observed in the first panel of Fig.\ref{fig5}. 
It is, thus, clear that the prolonged zero-field conditions in polar cap fields 
is linked to solar magnetic activity in the sunspot belt zone.  Furthermore, 
it is evident that the polar field strength in Cycle 24 during solar decline 
phase is much weaker than that in Cycle 23.  The good correlation between the 
polar cap field strength and the magnetic flux transport implies that the 
meridional flow speed could be much faster during the minimum of Cycle 24 
than the minimum in Cycle 23.  This, according to surface flux transport 
models, can produce weaker polar fields during the minimum of Cycle 24. 
The prolonged zero-field condition in the Northern hemisphere could 
be due to a faster meridional flow speed leading to more 
flux cancellation at the poles due to transport of more opposite 
magnetic flux from the sunspot belt region.
% 
%----------------------------- Begin Fig 6-----------------------------
\begin{center}
	\protect\begin{figure*}[!ht]
				\vspace{16.0cm}
				\includegraphics{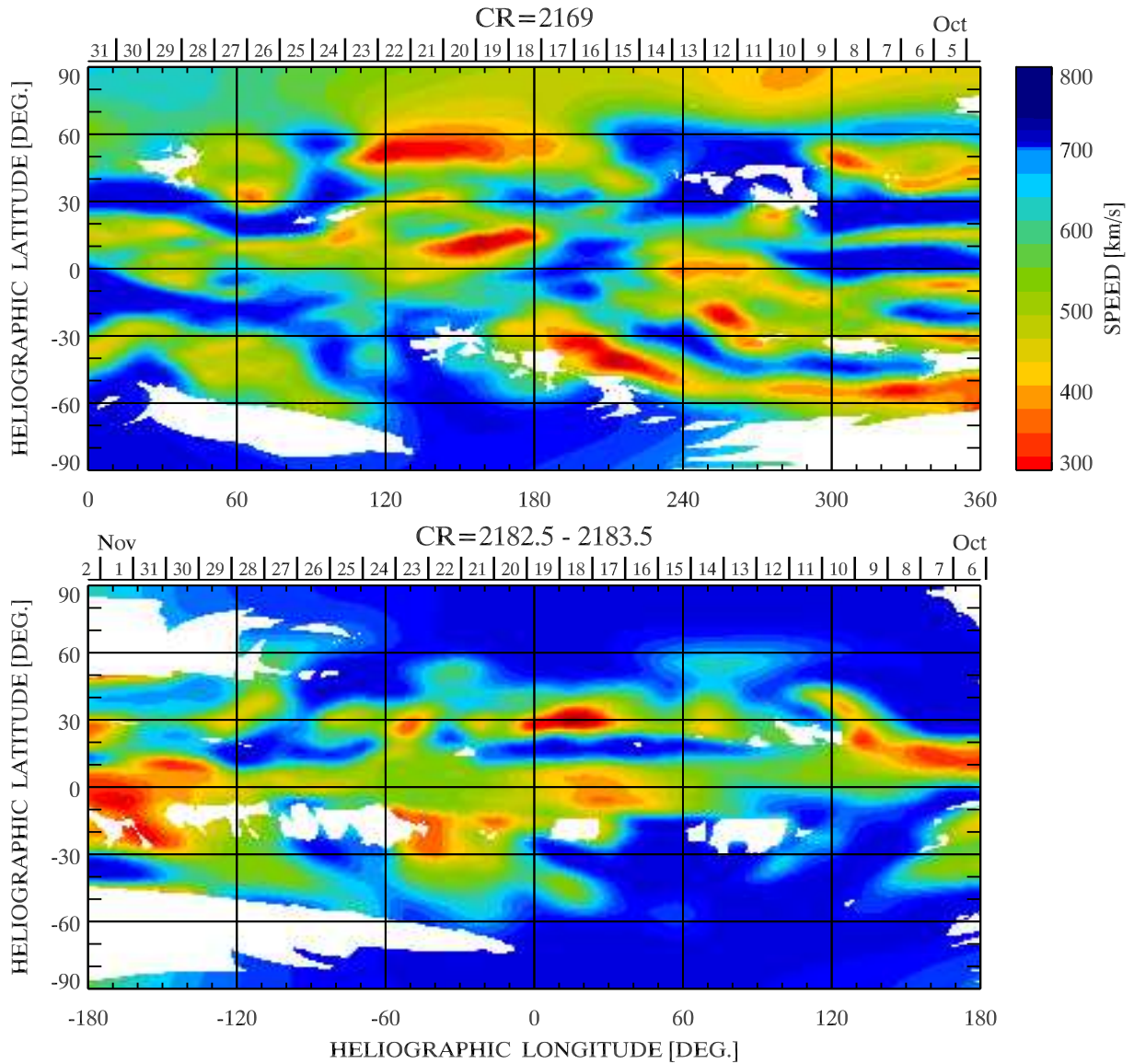}
		\caption{Synoptic velocity maps for Carrington rotation 
			2169 in October 2015 (upper panel) and October 2016 
			(lower panel). The large polar coronal hole in the 
			Southern hemisphere is apparent but its counterpart 
			in the Northern hemisphere has not formed in as of 
			October 2015 while it is fully developed in October 
			2016, a year later.}
		\label{fig6}
	\end{figure*}
\end{center}
%----------------------------------- End Fig6-----------------------
%
\subsection{Heliospheric signatures}

The heliospheric signatures of this extremely unusual polar field reversal pattern 
in cycle 24 can be seen in Figure \ref{fig6} which shows a synoptic map of solar wind 
velocities for Carrington rotation 2169 covering the period 04-31 October 2015 
(upper panel) and for Carrington Rotations 2182.5 to 2183.5 covering the period 
06 October 2016 to 02 November 2016(lower panel).  It is well known that polar coronal 
holes, which are the source regions of high speed solar wind \citep{KTR73, NoK76,Zir77}, 
begin to develop in both solar hemispheres after the field reversal takes place. 
Nearly four decades of solar wind observations from the three-station IPS facility 
at ISEE, Japan, have shown that polar coronal holes recover and become 
fully developed $\sim$ 2 years after the field reversal\citep{FuT16} and continue 
to expand and grow in area, as the minimum approaches.  \citet{FuT16} have in fact 
used an automated procedure to construct a database of 3335 coronal holes appearing 
between 1975 and 2015 and their study shows that the high latitude coronal holes recover 
in about 1.5 to 2 years after solar maximum.  The synoptic map in the lower panel of 
Fig. \ref{fig6} (06 Oct.--02 Nov. 2016) shows large, extended regions of 
high velocity solar wind outflows (in blue), or polar coronal holes, in the region 
poleward of $60^{\circ}$ in both hemispheres. It is clear from Fig.\ref{fig6} 
(upper panel: 04\,--\,31 Oct. 2015) that a well defined polar coronal hole 
was present in the Southern hemisphere by October 2015, $\sim$2 years after the field 
reversal in the south, that took place in Nov 2013, while it had not developed in 
the Northern hemisphere until a year later in $\sim$ Oct. 2016, as evident from 
Fig.\ref{fig6} (lower panel).  This is consistent with the asymmetric field reversal 
pattern seen in the Northern and Southern hemisphere in cycle 24, as shown in 
Fig.\ref{fig1}, which shows the completion of polarity reversal of fields in the 
Southern hemisphere in late 2013 and in the Northern hemisphere in late 2014.  The 
completion of polar reversal would have been followed by the development of the 
polar coronal hole in both the hemisphere. Since the completion of reversal 
occurred at different times in the two hemispheres, we see a difference in  
time of appearance of fully developed polar coronal holes in the two 
hemispheres.

\section{Summary and Conclusions} 
We have examined the polar field reversal process from solar cycles
21\,--\,24, spanning the period between Jan. 1975 and Dec. 2016, 
using magnetic field measurements from NSO/KP and the constructed 
magnetic butterfly diagrams which clearly depict the field reversal 
during each solar cycle.  Our study highlights the unusual nature
and the significant hemispheric asymmetry of the field reversal pattern 
in solar cycle 24 by examining high latitude solar magnetic fields, 
poleward of $55^{\circ}$ and $78^{\circ}$.  The current study shows that 
the field reversal in the Northern solar hemisphere was completed only by
November 2014 while the Southern hemisphere underwent a reversal in 
November 2013.  However, a recent study \citep{GoY16}, suggested that 
this process was completed only by the end of 2015, a full year 
later than our estimate from the present study.  It is to be noted that 
these authors used 17 GHz microwave images to show 
the absence of microwave brightness enhancement in the polar region 
during the prolonged zero-field polar field conditions in the Northern 
hemisphere following 2012 until late 2015.  The difference in completion of 
polar field reversal in the Northern hemisphere is understood from the fact 
that the 17GHz microwave emission is generally observed in the lower 
corona or at best in the chromospheric region, but not at the photospheric 
level.  However, our study of the polar reversal is at the photosphere and 
is therefore better in pinpointing the time of completion of the polar 
reversal in the solar Northern hemisphere.  Our study shows that the reversal 
occurred much earlier, in late 2014.  Confirmation of our results 
also came from an entirely different study aimed at examining the alignment 
of the sun's magnetic and rotational axis \citep{PMC15} using 
line-of-sight HMI magnetograms for a 5 year period starting from April 2010.  
Apart from finding a monthly oscillation at all solar latitudes which 
they attributed to a non-alignment in the solar magnetic and rotational axis, 
their data also indicated the time of occurrence of the solar polar field 
reversals to be in $\sim$ April 2014 for the Northern hemisphere and 
$\sim$ Feb. 2013 for the Southern Hemisphere [see Fig. 2 of \citet{PMC15}.]

The heliospheric signatures of this unusual polar reversal pattern, studied 
using synoptic maps of solar wind velocities, has shown the development of 
polar coronal holes in the Northern hemisphere as late as in October 2016. 
Since the heliospheric measurements of solar wind velocities are made in the 
corona at a height higher than that of the 17 GHz microwave emission, it is, 
thus, expected that the time of the polar reversal pattern could be later than 
the time of polar reversal  from the 17 GHz microwave brightness enhancement.

The polar reversal pattern, in cycle 24, in the solar Northern 
hemisphere was very unusual and probably unprecedented.  A field 
reversal or zero crossings of the magnetic field first occurred in June 2012.  
However, this was followed by a long period when the strength of the field did not
increase and remained nearly zero until November 2014, after which it showed a 
clear and unambiguous increase.  The Southern hemisphere, on the other hand, showed 
only a single unambiguous zero crossing or field reversal in November 2013, as can 
be seen in Fig.\ref{fig1}.   The hemispheric asymmetry of polar field 
reversal is well known and well discussed by \citet{SKa13} wherein, the 
authors attributed such hemispheric asymmetry to the asymmetry in solar 
activity in both the hemispheres.  Our study of the variation of toroidal 
magnetic fields showing a hemispheric asymmetry in solar activity in the sunspot 
belt regions lends support to the report by \citet{SKa13}.  The unusual pattern 
of field reversals in the Northern hemisphere in Cycle 24 can be 
attributed to multiple surges of solar activity after 2012 that carried the 
wrong magnetic flux to the solar north pole as seen by the latitudinal field 
profiles in the lower panel of Fig. \ref{fig3} in this study.  It is important 
to note here that a clean or unambiguous field reversal can occur when the 
magnetic flux transported to the poles has a polarity that is the opposite 
to that of the incumbent polarity.  It is to be noted that the present study 
very clearly shows the occurrence of multiple such surge activity in the Northern 
hemisphere, having the same polarity as the incumbent polarity in the north
during cycle 24. 

As opposed to an early study, carried out using 17 GHz microwave imaging 
observations \citep{GoY16}, that suggested that the zero field condition in 
the Northern hemisphere lasted until late 2015, our study of the latitudinal 
field profiles from the magnetic butterfly diagram as well as the unsigned 
polar field strength poleward of $78^{\circ}$ show that the zero field condition 
in the Northern hemisphere existed for nearly 2.5 years from $\sim$June 2012 
until $\sim$Nov. 2014.

Our study shows that the polar field strength during cycle 24 was 
comparatively weaker than in the previous cycle 23.  Also, the polar field strength 
in cycle 23 had been weaker than the earlier cycles 21--22 \citep{WRS09, JBG10}. 
Thus, it is clear that the polar field strength has been steady declining from 
cycle 21 to 24. \cite{JBG10} showed a very good correlation of polar field 
strength in the latitude range $78^{\circ}$\,--\,$90^{\circ}$ with the meridional 
flow speed, reported by \citep{HRi10} in cycle 23. In the present study, we 
also found a similar temporal behaviour of the polar field strength in the 
latitude range $78^{\circ}$\,--\,$90^{\circ}$ during cycle 24 implying that 
the polar field strength and the meridional flow speed are correlated. 
Since the polar field strength during cycle 24 is seen to be weaker in 
comparison to the earlier cycle 23, we, thus, believe that the meridional flow 
speed could have been faster in cycle 24 than cycle 23 leading to a weakening 
of the polar field strength in cycle 24.  Further, the considerably weaker 
polar field strength seen in the Northern hemisphere as compared to the 
Southern hemisphere implies that the meridional flow speed in cycle 24 is 
hemispherically asymmetric with a faster meridional flow in the north than 
the south.  As a result of the faster meridional flow in the north, surge 
activities of alternate polarities are noticed in the Northern hemisphere, which 
has, in turn, caused the prolonged zero-field conditions in the north by 
reducing the field strength due to multiple flux cancellations.
%  
%----------------------------- Begin Fig 7-----------------------------
\begin{center}
	\protect\begin{figure}[!ht]
	       \vspace{12.3cm}
	       \includegraphics{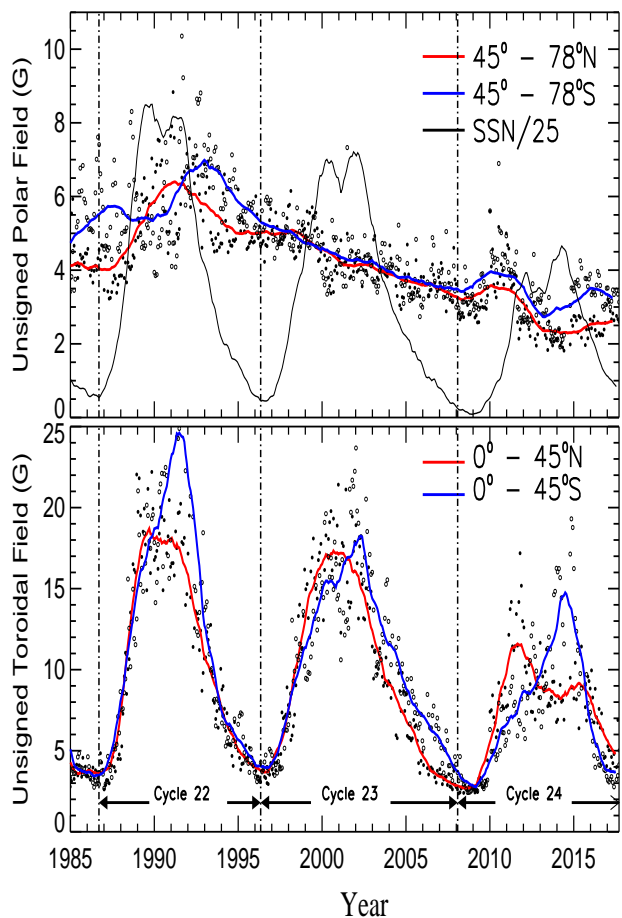}
		\caption{The upper panel shows, respectively, by the filled dots and 
		open circles, the unsigned values of solar polar fields for the Northern 
		and Southern hemispheres, obtained from NSO/KP magnetograms, in the 
		latitude ranges $45^{\circ}$\,--\,$78^{\circ}$, for solar cycles 22--24.  Also, 
		overplotted are the solid curves in red and blue showing the smoothed 
		variations of data points for the Northern and Southern hemispheres, respectively.  
		While the lower panel shows the same for the unsigned values of toroidal 
		fields in the latitude ranges $0^{\circ}$\,--\,$45^{\circ}$.  The smoothed total 
		sunspot number, scaled down by a factor of 25, is shown in solid black curve in 
		the upper panel. The vertical dotted lines demarcate the solar cycle periods.}
		\label{fig7}
	\end{figure}
\end{center}
%----------------------------------- End Fig 7-----------------------
%

It is known that the strength of the polar field can be used as precursor 
for predicting the strength of the upcoming cycle \citep{SPe93, Sch05}. 
The weaker polar field strength of cycle 24 implies a weaker cycle 25 in keeping   
with the flux transport dynamo model prediction of \citep{CCJ07} 
wherein, the authors used the axial dipole moments of the previous cycle 
to successfully predict the amplitude of the next cycle.  Using the 
long term variations in unsigned high-latitude ($45^{\circ}$\,--\,$78^{\circ}$) 
solar magnetic fields \citep{JaB15a}, a prediction of  $\sim$ 62 $\pm$12 has 
been made for the maximum sunspot number for solar cycle 25 in the 
old unmodified sunspot number scale indicating a  weaker cycle 25 
than cycle 24.  This study lends support to the prediction that the 
amplitude of the next cycle 25 will be weaker than the current cycle 24.  
A similar interpretation, that the cycle 25 will be comparatively 
weaker than the cycle 24, has also been obtained by \cite{UHa14}.

The upper panel of Figure \ref{fig7} shows by the filled dots and open circles, 
the unsigned values of solar polar fields for the Northern and Southern hemispheres, 
obtained from line-of-sight NSO/KP magnetograms, in the latitude ranges 
$45^{\circ}$\,--\,$78^{\circ}$.  Similarly, the lower panel shows the data for the latitude 
range $0^{\circ}$\,--\,$45^{\circ}$.  The data points shown start from 1985 (cycle 22) and 
cover the period until the end of 2017 in cycle 24.  Overplotted on the data points 
are solid curves in red (Northern) and blue (Southern) showing the smoothed variations 
of the data in the two hemispheres.  Earlier studies \citep{BiJ14,JaB15a} have 
reported a steady $\sim$20 year decline, starting from the mid-1990's to the end of 
2014 of solar photospheric fields in the latitude ranges $45^{\circ}$\,--\,$78^{\circ}$.  
It is evident from a careful examination of Fig. \ref{fig7} (upper panel) that the 
photospheric fields continuing to decline in the Northern hemisphere while the field 
strength in the Southern hemisphere has at least partially recovered and shown an 
increase since June 2014.  It can be seen from the lower panel of Fig.\ref{fig7} that 
the photospheric fields,at the end of 2017, have already approached the minimum in 
the Southern hemisphere while in the Northern hemisphere it is still declining and 
has yet to approach the minimum.  This asymmetry in the decline in both hemispheres 
and the continuation of photospheric field decline in the north explain the prolonged 
zero-field condition and delayed polar reversal of the Northern hemisphere.  Due to 
this the overall trend is still that of a decline and one would expect that this would 
continue at least until 2020, the expected minimum of the current cycle 24.  

The study of solar photospheric fields and a continuation of the 
study of solar polar field reversal process is, therefore, of utmost 
importance in understanding the sun, the solar dynamo process, the 
solar wind and space weather, more so because of the fact that such a 
situation on the sun, when solar photospheric fields have been steadily 
declining for nearly 25 years now and there is speculation that 
we are probably on the verge of a grand minimum akin to the Maunder 
minimum \citep{JaB15a,JaB15b,San16}, is unique and probably unprecedented 
since systematic solar observations began four centuries ago.

\begin{acknowledgements}
 We acknowledge the free data usage policy of NSO/KP, NSO/SOLIS, SoHO/MDI, 
 SDO/HMI and WDC-SILSO of Royal Observatory, Belgium facilities.  JP, DR 
 and KF acknowledge the ISEE International Collaborative Research Program 
 for support in executing this work.  SKB is supported by PIFI program of the 
 Chinese Academy of Sciences (Project Number: 2015PM066) and NSFC 
 (Grant no. 117550110422). 
\end{acknowledgements}
\bibliographystyle{aa}
%\bibliography{bibliography}  

\begin{thebibliography}{49}
\expandafter\ifx\csname natexlab\endcsname\relax\def\natexlab#1{#1}\fi

\bibitem[{{Altrock}(2011)}]{Alt11}
{Altrock}, R.~C. 2011, \solphys, 274, 251

\bibitem[{{Ananthakrishnan} {et~al.}(1995){Ananthakrishnan}, {Balasubramanian},
  \& {Janardhan}}]{ABJ95}
{Ananthakrishnan}, S., {Balasubramanian}, V., \& {Janardhan}, P. 1995, \ssr,
  72, 229

\bibitem[{{Babcock}(1959)}]{Bab59}
{Babcock}, H.~D. 1959, \apj, 130, 364

\bibitem[{{Babcock}(1961)}]{BaB61}
{Babcock}, H.~W. 1961, \apj, 133, 572

\bibitem[{{Benevolenskaya}(2004)}]{Ben04}
{Benevolenskaya}, E.~E. 2004, \aap, 428, L5

\bibitem[{{Bisoi} {et~al.}(2014{\natexlab{a}}){Bisoi}, {Janardhan},
  {Chakrabarty}, {Ananthakrishnan}, \& {Divekar}}]{BiJ14}
{Bisoi}, S.~K., {Janardhan}, P., {Chakrabarty}, D., {Ananthakrishnan}, S., \&
  {Divekar}, A. 2014{\natexlab{a}}, \solphys, 289, 41

\bibitem[{{Bisoi} {et~al.}(2014{\natexlab{b}}){Bisoi}, {Janardhan}, {Ingale},
  {Subramanian}, {Ananthakrishnan}, {Tokumaru}, \& {Fujiki}}]{BiJ14b}
{Bisoi}, S.~K., {Janardhan}, P., {Ingale}, M., {et~al.} 2014{\natexlab{b}},
  \apj, 793, 8

\bibitem[{{Choudhuri} {et~al.}(2007){Choudhuri}, {Chatterjee}, \&
  {Jiang}}]{CCJ07}
{Choudhuri}, A.~R., {Chatterjee}, P., \& {Jiang}, J. 2007, Physical Review
  Letters, 98, 131103

\bibitem[{{de Toma}(2011)}]{Tom11}
{de Toma}, G. 2011, \solphys, 274, 195

\bibitem[{{Dikpati} {et~al.}(2004){Dikpati}, {de Toma}, {Gilman}, {Arge}, \&
  {White}}]{DiG04}
{Dikpati}, M., {de Toma}, G., {Gilman}, P.~A., {Arge}, C.~N., \& {White}, O.~R.
  2004, \apj, 601, 1136

\bibitem[{{Dikpati} {et~al.}(2010){Dikpati}, {Gilman}, {de Toma}, \&
  {Ulrich}}]{DiG10}
{Dikpati}, M., {Gilman}, P.~A., {de Toma}, G., \& {Ulrich}, R.~K. 2010, \grl,
  37, 14107

\bibitem[{{Domingo} {et~al.}(1995){Domingo}, {Fleck}, \& {Poland}}]{DFP95}
{Domingo}, V., {Fleck}, B., \& {Poland}, A.~I. 1995, \solphys, 162, 1

\bibitem[{{Durrant} \& {Wilson}(2003)}]{DWi03}
{Durrant}, C.~J. \& {Wilson}, P.~R. 2003, \solphys, 214, 23

\bibitem[{{Fox} {et~al.}(1998){Fox}, {McIntosh}, \& {Wilson}}]{FMW98}
{Fox}, P., {McIntosh}, P., \& {Wilson}, P.~R. 1998, \solphys, 177, 375

\bibitem[{{Fujiki} {et~al.}(2016){Fujiki}, {Tokumaru}, {Hayashi}, {Satonaka},
  \& {Hakamada}}]{FuT16}
{Fujiki}, K., {Tokumaru}, M., {Hayashi}, K., {Satonaka}, D., \& {Hakamada}, K.
  2016, \apjl, 827, L41

\bibitem[{{Gopalswamy} {et~al.}(2016){Gopalswamy}, {Yashiro}, \&
  {Akiyama}}]{GoY16}
{Gopalswamy}, N., {Yashiro}, S., \& {Akiyama}, S. 2016, \apjl, 823, L15

\bibitem[{{Gopalswamy} {et~al.}(2012){Gopalswamy}, {Yashiro}, {M{\"a}kel{\"a}},
  {Michalek}, {Shibasaki}, \& {Hathaway}}]{GoY12}
{Gopalswamy}, N., {Yashiro}, S., {M{\"a}kel{\"a}}, P., {et~al.} 2012, \apjl,
  750, L42

\bibitem[{{Hathaway} \& {Rightmire}(2010)}]{HRi10}
{Hathaway}, D.~H. \& {Rightmire}, L. 2010, Science, 327, 1350

\bibitem[{{Hewish} {et~al.}(1964){Hewish}, {Scott}, \& {Wills}}]{HSW64}
{Hewish}, A., {Scott}, P.~F., \& {Wills}, D. 1964, \nat, 203, 1214

\bibitem[{{Howard}(1972)}]{How72}
{Howard}, R. 1972, \solphys, 25, 5

\bibitem[{{Janardhan} \& {Alurkar}(1993)}]{JAl93}
{Janardhan}, P. \& {Alurkar}, S.~K. 1993, \aap, 269, 119

\bibitem[{{Janardhan} {et~al.}(1996){Janardhan}, {Balasubramanian},
  {Ananthakrishnan}, {Dryer}, {Bhatnagar}, \& {McIntosh}}]{JaB96}
{Janardhan}, P., {Balasubramanian}, V., {Ananthakrishnan}, S., {et~al.} 1996,
  \solphys, 166, 379

\bibitem[{{Janardhan} {et~al.}(2015{\natexlab{a}}){Janardhan}, {Bisoi},
  {Ananthakrishnan}, {Sridharan}, \& {Jose}}]{JaB15b}
{Janardhan}, P., {Bisoi}, S.~K., {Ananthakrishnan}, S., {Sridharan}, R., \&
  {Jose}, L. 2015{\natexlab{a}}, Sun and Geosphere, 10, 147

\bibitem[{{Janardhan} {et~al.}(2011){Janardhan}, {Bisoi}, {Ananthakrishnan},
  {Tokumaru}, \& {Fujiki}}]{JaB11}
{Janardhan}, P., {Bisoi}, S.~K., {Ananthakrishnan}, S., {Tokumaru}, M., \&
  {Fujiki}, K. 2011, \grl, 38, L20108

\bibitem[{{Janardhan} {et~al.}(2015{\natexlab{b}}){Janardhan}, {Bisoi},
  {Ananthakrishnan}, {Tokumaru}, {Fujiki}, {Jose}, \& {Sridharan}}]{JaB15a}
{Janardhan}, P., {Bisoi}, S.~K., {Ananthakrishnan}, S., {et~al.}
  2015{\natexlab{b}}, \jgr, 120, 5306

\bibitem[{{Janardhan} {et~al.}(2010){Janardhan}, {Bisoi}, \& {Gosain}}]{JBG10}
{Janardhan}, P., {Bisoi}, S.~K., \& {Gosain}, S. 2010, \solphys, 267, 267

\bibitem[{{Jian} {et~al.}(2011){Jian}, {Russell}, \& {Luhmann}}]{JRL11}
{Jian}, L.~K., {Russell}, C.~T., \& {Luhmann}, J.~G. 2011, \solphys, 274, 321

\bibitem[{{Karna} {et~al.}(2014){Karna}, {Hess Webber}, \& {Pesnell}}]{KHP14}
{Karna}, N., {Hess Webber}, S.~A., \& {Pesnell}, W.~D. 2014, \solphys, 289,
  3381

\bibitem[{{Kirk} {et~al.}(2009){Kirk}, {Pesnell}, {Young}, \& {Hess
  Webber}}]{KiP09}
{Kirk}, M.~S., {Pesnell}, W.~D., {Young}, C.~A., \& {Hess Webber}, S.~A. 2009,
  \solphys, 257, 99

\bibitem[{{Kojima} {et~al.}(1998){Kojima}, {Tokumaru}, {Watanabe}, {Yokobe},
  {Asai}, {Jackson}, \& {Hick}}]{KoT98}
{Kojima}, M., {Tokumaru}, M., {Watanabe}, H., {et~al.} 1998, \jgr, 103, 1981

\bibitem[{{Krieger} {et~al.}(1973){Krieger}, {Timothy}, \& {Roelof}}]{KTR73}
{Krieger}, A.~S., {Timothy}, A.~F., \& {Roelof}, E.~C. 1973, \solphys, 29, 505

\bibitem[{{Leighton}(1969)}]{Lei69}
{Leighton}, R.~B. 1969, \apj, 156, 1

\bibitem[{{Lemen} {et~al.}(2012){Lemen}, {Title}, {Akin}, {Boerner}, {Chou},
  {Drake}, {Duncan}, {Edwards}, {Friedlaender}, {Heyman}, {Hurlburt}, {Katz},
  {Kushner}, {Levay}, {Lindgren}, {Mathur}, {McFeaters}, {Mitchell}, {Rehse},
  {Schrijver}, {Springer}, {Stern}, {Tarbell}, {Wuelser}, {Wolfson}, {Yanari},
  {Bookbinder}, {Cheimets}, {Caldwell}, {Deluca}, {Gates}, {Golub}, {Park},
  {Podgorski}, {Bush}, {Scherrer}, {Gummin}, {Smith}, {Auker}, {Jerram},
  {Pool}, {Soufli}, {Windt}, {Beardsley}, {Clapp}, {Lang}, \&
  {Waltham}}]{LeT12}
{Lemen}, J.~R., {Title}, A.~M., {Akin}, D.~J., {et~al.} 2012, \solphys, 275, 17

\bibitem[{{Makarov} {et~al.}(1983){Makarov}, {Fatianov}, \&
  {Sivaraman}}]{MFS83}
{Makarov}, V.~I., {Fatianov}, M.~P., \& {Sivaraman}, K.~R. 1983, \solphys, 85,
  215

\bibitem[{{Moran} {et~al.}(2000){Moran}, {Ananthakrishnan}, {Balasubramanian},
  {Breen}, {Canals}, {Fallows}, {Janardhan}, {Tokumaru}, \& {Williams}}]{MoA00}
{Moran}, P.~J., {Ananthakrishnan}, S., {Balasubramanian}, V., {et~al.} 2000,
  Annales Geophysicae, 18, 1003

\bibitem[{{Mu{\~n}oz-Jaramillo} {et~al.}(2012){Mu{\~n}oz-Jaramillo}, {Sheeley},
  {Zhang}, \& {DeLuca}}]{MuS12}
{Mu{\~n}oz-Jaramillo}, A., {Sheeley}, N.~R., {Zhang}, J., \& {DeLuca}, E.~E.
  2012, \apj, 753, 146

\bibitem[{{Nolte} {et~al.}(1976){Nolte}, {Krieger}, {Timothy}, {Gold},
  {Roelof}, {Vaiana}, {Lazarus}, {Sullivan}, \& {McIntosh}}]{NoK76}
{Nolte}, J.~T., {Krieger}, A.~S., {Timothy}, A.~F., {et~al.} 1976, \solphys,
  46, 303

\bibitem[{{Pastor Yabar} {et~al.}(2015){Pastor Yabar}, {Mart{\'{\i}}nez
  Gonz{\'a}lez}, \& {Collados}}]{PMC15}
{Pastor Yabar}, A., {Mart{\'{\i}}nez Gonz{\'a}lez}, M.~J., \& {Collados}, M.
  2015, \mnras, 453, L69

\bibitem[{{Petrie}(2012)}]{Pet12}
{Petrie}, G.~J.~D. 2012, \solphys, 281, 577

\bibitem[{{S{\'a}nchez-Sesma}(2016)}]{San16}
{S{\'a}nchez-Sesma}, J. 2016, Earth System Dynamics, 7, 583

\bibitem[{{Schatten}(2005)}]{Sch05}
{Schatten}, K. 2005, \grl, 32, 21106

\bibitem[{{Schatten} \& {Pesnell}(1993)}]{SPe93}
{Schatten}, K.~H. \& {Pesnell}, W.~D. 1993, \grl, 20, 2275

\bibitem[{{Sun} {et~al.}(2015){Sun}, {Hoeksema}, {Liu}, \& {Zhao}}]{SuH15}
{Sun}, X., {Hoeksema}, J.~T., {Liu}, Y., \& {Zhao}, J. 2015, \apj, 798, 114

\bibitem[{{Svalgaard} \& {Kamide}(2013)}]{SKa13}
{Svalgaard}, L. \& {Kamide}, Y. 2013, \apj, 763, 23

\bibitem[{{Upton} \& {Hathaway}(2014)}]{UHa14}
{Upton}, L. \& {Hathaway}, D.~H. 2014, \apj, 780, 5

\bibitem[{{Wang} {et~al.}(1989){Wang}, {Nash}, \& {Sheeley}}]{WNS89}
{Wang}, Y.-M., {Nash}, A.~G., \& {Sheeley}, Jr., N.~R. 1989, \apj, 347, 529

\bibitem[{{Wang} {et~al.}(2009){Wang}, {Robbrecht}, \& {Sheeley}}]{WRS09}
{Wang}, Y.-M., {Robbrecht}, E., \& {Sheeley}, Jr., N.~R. 2009, \apj, 707, 1372

\bibitem[{{Webb} {et~al.}(1984){Webb}, {Davis}, \& {McIntosh}}]{WDM84}
{Webb}, D.~F., {Davis}, J.~M., \& {McIntosh}, P.~S. 1984, \solphys, 92, 109

\bibitem[{{Zirker}(1977)}]{Zir77}
{Zirker}, J.~B. 1977, Reviews of Geophysics and Space Physics, 15, 257

\end{thebibliography}

%
\end{document}